\documentclass[prl,twocolumn,showpacs,showkeys,secnumarabic,amssymb,amsmath,unsortedaddress]{revtex4-1}

\usepackage{amssymb,amsmath}
\usepackage{color}
\usepackage{graphics}
\usepackage{graphicx}
\usepackage{bbold}
\usepackage{ulem}

\def\beq{\begin{equation}}
\def\eeq{\end{equation}}
\def\bea{\begin{eqnarray}}
\def\eea{\end{eqnarray}}

\begin{document}

\title{Distribution of quantum coherence in multipartite systems}

\author{Chandrashekar Radhakrishnan}
\affiliation{New York University, 1555 Century Avenue, Pudong, Shanghai 200122, China}
\affiliation{NYU-ECNU Institute of Physics at NYU Shanghai, 3663 Zhongshan Road North, Shanghai 200062, China}

\author{Manikandan Parthasarathy}
\author{Segar Jambulingam}
\affiliation{Department of Physics, Ramakrishna Mission Vivekananda College, Mylapore, Chennai 600004, India} 

\author{Tim Byrnes}
\affiliation{New York University, 1555 Century Avenue, Pudong, Shanghai 200122, China} 
\affiliation{NYU-ECNU Institute of Physics at NYU Shanghai, 3663 Zhongshan Road North, Shanghai 200062, China}
\affiliation{Department of Physics, New York University, New York 10003, USA}  
\affiliation{National Institute of Informatics, 2-1-2 Hitotsubashi, Chiyoda-ku, Tokyo 101-8430, Japan}

\begin{abstract}
The distribution of coherence in multipartite systems is examined. We use a new coherence measure with entropic nature and metric properties, based on the quantum Jensen-Shannon divergence.  The metric property allows for the coherence to be decomposed into various contributions, which arise from local and intrinsic coherences.  We find that there are trade-off relations between the various contributions of coherence, as a function of parameters of the quantum state. In bipartite systems the coherence resides on individual sites or distributed among the sites, which contribute in a complementary way. In more complex systems, the characteristics of the coherence can display more subtle changes with respect to the parameters of the quantum state. In the case of the $ XXZ $ Heisenberg model, the coherence changes from a monogamous to a polygamous nature.  This allows us to define the shareability of  coherence, leading to monogamy relations for coherence.  
\end{abstract}


\date{\today}

\maketitle
The concept of wave particle duality introduced the importance of quantum coherence in physical phenomena such as low temperature 
thermodynamics \cite{vs15}, quantum thermodynamics \cite{ja14,ml15,lo15}, nanoscale physics \cite{ok11}, biological systems 
\cite{ps07,er14}, and is one of the most basic aspects of quantum information science \cite{mn00}.  For this reason, understanding quantum coherence has a long history and is of fundamental importance to many fields.  In quantum optics \cite{rg63,es63}, the approach has been typically to examine quantities such as phase space distributions and higher order correlation functions \cite{ms97}. While this method distinguishes between quantum and classical coherence, it does not quantify coherence in a rigorous sense.
More recently, a procedure to quantify coherence using methods of quantum information science was developed \cite{tb14,dg14,dpp15,as15}. 
In the seminal work of Ref. \cite{tb14}, basic quantities such as incoherent states, incoherent operations, maximally coherent states were 
defined and the set of properties a functional should satisfy to be considered as a coherence measure were listed.

One fundamental task that is desirable is to pinpoint what part of a quantum system is responsible for any coherence that is present.
  To understand the possibilities, let us consider a two qubit system as an example. Coherence is a basis-dependent quantity \cite{as15,yy15}, and the reference incoherent states are chosen as $ |0 \rangle, | 1 \rangle $.   We can consider then two types of states which possess coherence,
$ ( |0 \rangle - |1 \rangle )( |0 \rangle - |1 \rangle ) $ and $ |0 \rangle |0 \rangle - | 1 \rangle | 1 \rangle $.  In the former, the coherence lies on each qubit, while the latter has a kind of collective coherence, i.e. entanglement. An interesting aspect of this is that the types of coherence are complementary to each other -- an increase in one type leads to a corresponding decrease in the other. In order to have maximum coherence on a particular qubit, it is optimal to create a superposition on each one, which excludes entanglement.  On the other hand, for the Bell state, tracing out one of the qubits leaves a completely mixed (incoherent) state on the other qubit.

\begin{figure}
\includegraphics[width=\columnwidth]{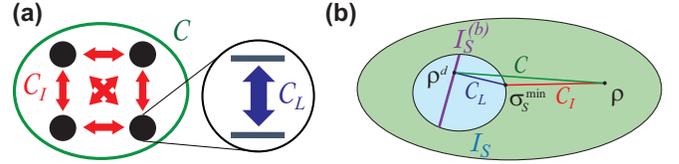}
\caption{\label{fig1}
Quantum coherence in multipartite systems.  (a)  The total coherence $ \mathcal{C} $ has contributions from local 
coherence $ \mathcal{C}_L $ on subsystems and collective coherence $ \mathcal{C}_I $.  (b) Definitions of various coherences
according to the distance between states. $ \mathcal{I}_S $ is the set of separable states, while $ \mathcal{I}^{(b)}_S $ is the set of separable states in a fixed basis $ b $. $ \rho_d $ is the solution of (\ref{coherencedef}) and $ \sigma_S^{\min} $ is the solution of (\ref{intrinsiccoherence}). }
\end{figure}

This complementary behavior is reminiscent of another quantum feature, monogamy of entanglement, 
which has attracted a lot of attention recently \cite{vc00,mk04,to06,ga06,pk14}. Monogamy is a concept related to the shareability of entanglement between different constituents in a multipartite system. For example, in a tripartite system, if Alice and Bob have a maximally entangled state then this rules out entanglement to Charlie. The monogamy relation for three qubits was introduced in Ref. \cite{vc00}, has also been generalized to multipartite systems \cite{to06}.  Both these examples illustrate the trade-off nature of quantum mechanical features, where increasing one imposes restrictions on the other. 
Another fundamental question which this raises is the relationship between coherence and entanglement
\cite{as15,yy15}. 
The framework outlined in Ref. \cite{tb14} closely followed the format of entanglement quantification
developed in \cite{vv97,vvp97,vv98}. While entanglement is clearly a form of coherence, the converse is not necessarily 
not true.  In this paper we explore the question of how we can quantify various types of coherence, and examine their trade-off relations within a multipartite system.  By understanding the distribution of coherence in a multipartite system, this leads us to find the relation between concepts such as coherence, entanglement, and monogamy. 

One of the tools that we will use in this study is a coherence measure which has both entropic and geometric properties.
In Ref. \cite{tb14}, two different functionals, one based on the relative entropy and the other based on the $\ell_{1}$-norm were found to 
satisfy the necessary properties as a coherence measure.  Of these, the former is an entropic measure while the other is a geometric measure which can be used as a formal distance measure. Any measure ${\mathcal D}$ is considered as a formal distance over the set $X$ if $\forall$ $\rho,\sigma \in X$ it satisfies the following properties: (i) ${\mathcal D} (\rho,\sigma) > 0$ for $\rho \neq \sigma$ and ${\mathcal D} (\rho,\rho) = 0$, (ii) ${\mathcal D} (\rho,\sigma) = {\mathcal D}(\sigma,\rho)$ (symmetry).  If ${\mathcal D}$ satisfies (iii) ${\mathcal D}(\rho,\sigma) + {\mathcal D}(\sigma,\tau) \geq {\mathcal D}(\rho,\tau)$ (the triangle inequality) in addition to the properties given above, then $\mathcal{D}$ is a metric for the space $X$.  The relative entropy $S(\rho \| \sigma) \equiv  \text{Tr} \rho \log (\rho/\sigma)$ is not a distance since it is asymmetric and further it is well defined only when the support of $\sigma$ is equal to or larger than that of $\rho$.  Towards this 
end we introduce here an alternative, the quantum version of the Jensen-Shannon divergence (QJSD):
\begin{align}
\mathcal{J}(\rho,\sigma) = \frac{1}{2} [S(\rho\|(\rho+\sigma)/2) + S(\sigma\|(\rho+\sigma)/2)].
\end{align}
The QJSD is known to be a distance measure, be bounded $0 \leq \mathcal{J} \leq 1$, and is well defined
irrespective of the nature of the support of $\rho$ and $\sigma$ \cite{jp09,ap05,pl08}. 
The QJSD does not obey the triangle inequality, but its square root obeys it for all 
pure states. In the case of mixed states there is no general proof of the triangle 
inequality, but numerical studies up to five qubits \cite{pl08} strongly indicate its validity.

{\it Quantum coherence trade-offs.} 
The quantum coherence is defined as \cite{tb14}
\begin{align}
{\mathcal C} (\rho)  & \equiv \underset{\sigma \in \mathcal{I}^{(b)}}{\hbox{min}} {\mathcal D}  (\rho,\sigma),
\label{coherencedef}
\end{align}
where $ {\mathcal D} $ is a distance measure and $\mathcal{I}^{(b)}$ are the set of incoherent states in a particular basis $ b $. The functional ${\mathcal C}$ is a quantum coherence measure if it obeys the properties \cite{tb14}: (i) ${\mathcal C} (\rho)  \geq 0$ and ${\mathcal C}(\rho) = 0$ iff $\rho \in \mathcal{I}^{(b)} $; (ii) ${\mathcal C} (\rho)$ is invariant under unitary transformations; (iii) ${\mathcal C} (\rho)$ is monotonic under a ICPTP (incoherent completely positive and trace preserving map); (iv) ${\mathcal C} (\rho)$ monotonic under selective incoherent measurements on average; and (v) ${\mathcal C}(\rho)$ non-increasing under mixing of quantum states (convexity). 

Eq. (\ref{coherencedef}) states that the amount of coherence in a given state is the distance to the closest incoherent state.  This definition clearly depends on what we deem to be an incoherent state, and is responsible for the basis-dependent nature of $ \mathcal C $.  Most generally, one may assume a form for an incoherent state $ \sigma = \sum_{k} p_{k} |b_k \rangle \langle b_k| $ 
where the $\{|b_k \rangle \}$ are a fixed particular basis choice $ b $, and $ p_{k} $ are probabilities.  Without the constraint of the fixed basis, it is always possible to write $ \sigma = \rho $ by taking $ | b_k\rangle $ to be eigenvectors of $ \rho $, which immediately gives $ {\mathcal C} = 0 $. In this paper we are interested in how the overall coherence is distributed in a multipartite system. For this reason it will be most interesting to choose a local basis choice
\begin{equation}
\sigma = \sum_{k} p_{k} \tau_{k,1}^{(b)} \otimes \cdots \otimes \tau_{k,N}^{(b)}, 
\label{separablestates} 
\end{equation}
where $\tau_{k,n}^{(b)}$ is the incoherent state on the subsystem $n$ i.e., 
$\tau_{k,n}^{(b)} = \sum_{k} p_{k,n} |b_{k,n} \rangle \langle b_{k,n}|$. The set of states that are separable and in a basis $ b $ are called $\mathcal{I}^{(b)}_S$.  

This gives a natural way to study various coherence contributions within a multipartite system.  As discussed above,  we can distinguish between coherence that is localized on the subsystems $ n $, and and collective coherence which cannot be attributed to particular subsystems (see Fig. \ref{fig1}(a)).  To remove the contribution from the subsystems, we may relax the basis constraint $ b $ and minimize over the set of states (\ref{separablestates}).  This contribution is independent of the basis choice, and is the coherence which is intrinsic within the system.  We thus define the intrinsic coherence 
\begin{align}
{\mathcal C}_I (\rho)  \equiv \underset{\sigma_S \in \mathcal{I}_S}{\hbox{min}} {\mathcal D}  (\rho,\sigma_S),
\label{intrinsiccoherence}
\end{align}
where $ \mathcal{I}_S $ is the set of states of the form as given in (\ref{separablestates}), but is not necessarily in the basis $ b $. 
 Thus the only constraint here is the general form of the basis, that it is separable, but the particular basis is not specified. Eq. (\ref{intrinsiccoherence}) is in fact equal to the entanglement, which is reasonable from the point of view that entanglement must contribute to coherence \cite{as15}. The remaining contribution then originates from coherence that exists on the subsystems, and we can write the local coherence as
\begin{align}
{\mathcal C}_L (\rho) \equiv {\mathcal D}  (\sigma_S^{\min} , \rho^d ),
\label{localcoherence}
\end{align}
where $ \sigma_S^{\min} $ and $ \rho^d $ are the minimum solutions of (\ref{intrinsiccoherence}) and (\ref{coherencedef})  respectively, and are implicit functions of $ \rho $.

We may visualize the two different contributions according to Fig. \ref{fig1}(b). According to the metric properties of $ \mathcal{D} $, and the triangle inequality we immediately see that
\begin{align}
{\mathcal C} \le  {\mathcal C}_L + {\mathcal C}_I  ,
\label{localintrinsic}
\end{align}
For a product state $\sigma_S^{\min}$, the coherence measure is subadditive which leads to 
$ {\mathcal C}_L \le \sum_{n} {\mathcal C}_{L,n} $. We thus have
\begin{align}
{\mathcal C} \le  \sum_{n=1}^N {\mathcal C}_{L,n}  + {\mathcal C}_I,
\label{localintrinsic2}
\end{align}
where $ {\mathcal C}_{L,n} $ is the coherence on each subsystem $ n $ separately. 

\begin{figure}
\includegraphics[width=\columnwidth]{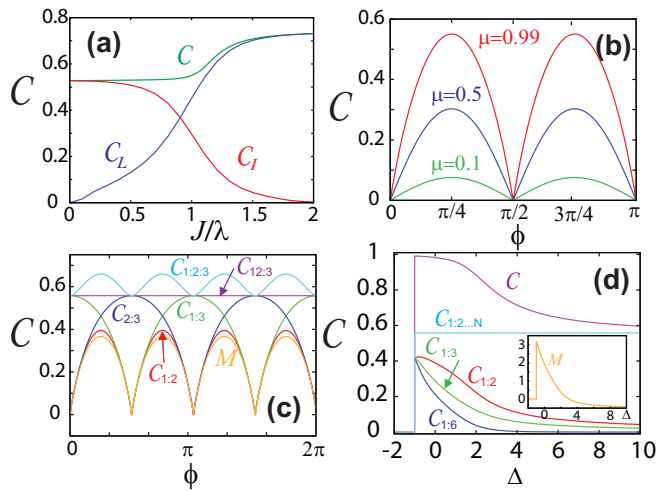}
\caption{\label{fig2}
Coherence as measured by the quantum Jensen-Shannon divergence for various states.  Coherence of (a) the $ N = 2 $ site Ising model with $ \epsilon=0.2 $; (b) Werner GHZ state; (c) W state with $ \theta = \pi/4 $; (d) The $ N = 10 $ site $XXZ$ Heisenberg model ground state with $ J= 1 $. Inset: Monogamy for the $XXZ $ Heisenberg model as defined in  (\ref{q-def}).}
\end{figure}

An illustrative example of the coherence decomposition is given by the ground state of the $N=2$ Ising model described by the Hamiltonian 
\begin{align}
H = \lambda \sigma_{1}^{x} \sigma_{2}^{x}  + J (\sigma_{1}^{x} + \sigma_{2}^{x}) + \epsilon \lambda  (\sigma_{1}^{z} + \sigma_{2}^{z}) ,
\end{align}
where $J,\lambda$ coupling parameters and $ \epsilon $ is a small symmetry breaking term.  The numerically estimated values of 
${\mathcal C}_L$, ${\mathcal C}_I$, and ${\mathcal C}$ are given in Fig. \ref{fig2}(a), where we use the square root of the Jensen-Shannon divergence as our distance measure
\begin{align}
{\mathcal D}  (\rho,\sigma)= \sqrt{\mathcal{J}(\rho,\sigma)} = 
\sqrt{S\left(\frac{\rho+\sigma}{2}\right) - \frac{S(\rho)}{2} - \frac{S(\sigma)}{2}}, \nonumber
\end{align}
where $ S( \rho) = - \text{Tr} \rho \log \rho $ is the von Neumann entropy. Taking the $ \{ |0 \rangle_n, |1 \rangle_n \} $ basis as the reference state (this will be the case throughout this paper), we see that there is a crossover between coherence contributions from intrinsic when $ J \ll \lambda $, to local as $ J \gg \lambda $.  This is due to the fact that for $ J = 0, \epsilon \rightarrow 0 $ the ground state approaches a Bell state $ | 00 \rangle - | 11 \rangle $, and for $ \lambda = 0 $ the ground state is $ (|0\rangle-|1 \rangle  ) (|0\rangle-|1 \rangle ) $, with intermediate $ J/\lambda $ giving an interpolation to these limits.  
The total coherence is less than the sum of the local and intrinsic contributions, following (\ref{localintrinsic}).

{\it Multipartite coherence.}
The bipartite case studied above is the simplest case of more general trade-off relations in multipartite systems. One of the fundamental properties we investigate is the shareability of coherence between subsystems. For example, in a tripartite system $ \rho_{123} $ we may decompose the coherence using (\ref{localintrinsic}) according to 
\begin{align}
\mathcal{C}_{123} \leq \mathcal{C}_1 + \mathcal{C}_2 + \mathcal{C}_3  + \mathcal{C}_{1:2:3} .
\label{tripartite1}
\end{align}
where we have introduced a shorthand for the local coherence on subsystem $ n $ as $ \mathcal{C}_n = \mathcal{C}_{L,n} (\rho_n) $, 
and $\rho_n $ is the reduced density matrix. For the product states $\sigma_S^{\min}$ the Eqn. (\ref{tripartite1}) holds 
exactly. 
The intrinsic coherence $ \mathcal{C}_{1:2:3} = \mathcal{C}_I (\rho_{123}) $ is minimized over the set of separable states on the tripartite system.  We note that as $ \mathcal{C}_{1:2:3} $ is an intrinsic coherence, it does not contain any coherence located {\it on} the sites, but contains all coherences {\it between} the sites. 

We can decompose a tripartite system in a bipartite fashion, leading to the relation
\begin{align}
\mathcal{C}_{123} \leq \mathcal{C}_1 + \mathcal{C}_2 + \mathcal{C}_3 + \mathcal{C}_{2:3}+ \mathcal{C}_{1:23},
\label{tripartite2}
\end{align}
where we first find the intrinsic coherence between 2 and 3, then we estimate the intrinsic 
coherence between 1 and the bipartite subsystem 23.  Similar decompositions can be carried out with respect to the 
bipartitions 2:13 and 3:12 as well.  From (\ref{tripartite1}) and (\ref{tripartite2}) and the other possible bipartitions
suggested above we may deduce that
\begin{align}
\mathcal{C}_{1:2:3} & \simeq \mathcal{C}_{2:3}+ \mathcal{C}_{1:23}  \simeq  \mathcal{C}_{1:2}+ \mathcal{C}_{12:3}
\simeq  \mathcal{C}_{1:3}+ \mathcal{C}_{13:2} .
\label{tribipartite}
\end{align}

To illustrate the various contributions, first let us consider the mixed GHZ states defined as
$
\rho_{\text{GHZ}} = \frac{1 - \mu}{8} {\hat{\mathbb{1}}} + \mu \; |\text{GHZ} \rangle \langle \text{GHZ}  |
$
with $ |\text{GHZ}  \rangle = \cos \phi |000 \rangle + \sin \phi |111 \rangle $, $\phi \in [0, 2\pi)$, and $0\le  \mu \le 1$.
The coherence is plotted in Fig. \ref{fig2}(b).  For this class of states we find that the various contributions due to one and two sites are always zero: $ \mathcal{C}_{n} = \mathcal{C}_{m:n}=0 $.  This means that the only coherence contributions originates from the intrinsic coherence where all three sites are involved.  The total coherence is thus identical to the tripartite coherence $ \mathcal{C} = \mathcal{C}_{1:2:3} $, which is verified numerically.  It is also equal to the bipartitioned intrinsic coherence $ \mathcal{C} =  \mathcal{C}_{l:mn} $, where $ l,m,n $ are all permutations of the sites.  This verifies the relation (\ref{tribipartite}) for this class of states. 

In contrast to the GHZ state where there is only one coherence contribution, the W states have a trade-off relation similar to that seen in the transverse Ising model.  These are defined
$
|\text{W} \rangle = \sin \theta \cos \phi |100 \rangle + \sin \theta \sin \phi |010 \rangle + \cos \theta |001 \rangle
$
with $0 \leq \phi < 2\pi$ and $0 \leq \theta \leq \pi$.  The GHZ and W states are two classes of states which are unrelated under local operations and classical communications.  From Fig. \ref{fig2}(c) we see that the calculated coherence can be attributed to several contributions.  Firstly the coherence $ \mathcal{C}_{12:3} $ is always constant as the state for the choice $ \theta = \pi/4 $ can be written as $|\text{W} \rangle = [ (\cos \phi |10 \rangle + \sin \phi  |01 \rangle )|0 \rangle + |00 \rangle| 1 \rangle ]/\sqrt{2} $, thus there is always intrinsic coherence between the bipartition of sites 12 and 3.  The coherences $ \mathcal{C}_{1:3} $ and $ \mathcal{C}_{2:3} $ show complementary behavior as the system oscillates between a Bell state between sites 13 ($\phi=n \pi$) and 23 ($\phi=(n+1/2)\pi$), with the remaining site being decoupled.  There is coherence between the sites 12 between these two extrema, giving $ \mathcal{C}_{1:2} $ with twice the oscillatory frequency. 

The same ideas can be equally applied to more complex multipartite systems.  The various coherence contributions can be used to understand the nature of the quantum states in quantum many-body systems. We illustrate this by analyzing the one-dimensional Heisenberg $XXZ$ model, one of the fundamental models in magnetism.  The Hamiltonian of this model is 
\begin{align}
H = J  \sum_{n} (\sigma^{x}_{n} \sigma^{x}_{n+1} + \sigma^{y}_{n} \sigma^{y}_{n+1} + \Delta \sigma^{z}_{n} \sigma^{z}_{n+1}),
\end{align}
where $J$ is the nearest neighbor spin coupling and $\Delta$ is the anisotropy parameter.  For an antiferromagnetic coupling $ J> 0 $, the system has a phase transition from the ferromagnetic axial regime to the antiferromagnetic planar regime at $\Delta =-1$.  Using exact diagonalization techniques we estimate various types of coherence as shown in Fig. \ref{fig2}(d). In the ferromagnetic phase with $ \Delta <-1 $, all coherences vanish due to spontaneous symmetry breaking selecting a unique ferromagnetic ground state with all spins aligned in the $ \sigma^{z} $ basis. In the opposite limit $ \Delta \gg 1 $, the state is a superposition of N\'{e}el states, due to the two-fold degeneracy of these states: $ (|0101 \dots 01 \rangle + |1010 \dots 10 \rangle)/\sqrt{2} $.  The coherence thus approaches the Bell state value $ \mathcal{C} = \mathcal{C}_{1:2 \dots N} \approx 0.56 $, with all other coherence contributions vanishing. Due to the spin flip symmetry, coherence on each site is always zero $ \mathcal{C}_{n} = 0 $, and $ \mathcal{C}_{1:2 \dots N}   $ can always be written in a Bell state form, resulting in a constant value.  The coherence contributions between two sites decrease with distance as expected $ \mathcal{C}_{1:n} $, due to the reduced correlations between these sites.  Interestingly, at $ \Delta = -1 $ the two-site correlations all converge to the same value, which we attribute to the fact that this is close to the antiferromagnetic-ferromagnetic phase transition, which has the effect of increasing the overall coherence in the system.

{\it Monogamy of coherence.}
From our coherence decompositions, we arrive naturally at the notion of monogamy of coherence. In a tripartite system, if subsystems 2 and 3 are maximally coherent with respect each other, this limits on the amount of coherence that subsystem 1 has with 2 and 3.  This is immediately evident from Eq. (\ref{tripartite2}), where the coherence is decomposed into these two contributions.  If subsystem 3 is coherently connected to 1 and 2 then the tripartite system is described to be polygamous, and otherwise is monogamous. The coherence monogamy relations may be identified from (\ref{tribipartite}), where we observe that the tripartite coherence $ \mathcal{C}_{1:2:3} $ can be decomposed into several bipartite coherences. The genuine tripartite coherence can be estimated by subtracting pairwise bipartite terms giving
\begin{align}
\mathcal{C}_{1:2:3} - & \mathcal{C}_{1:2} -\mathcal{C}_{2:3} - \mathcal{C}_{1:3}  \simeq  \mathcal{C}_{1:23} -\mathcal{C}_{1:2}  -\mathcal{C}_{1:3} .
\end{align}
For a multipartite system the monogamy inequality reads  $ \mathcal{C}_{1:2\dots N} \geq \sum_{n=2}^{N} \mathcal{C}_{1:n} $.  
Thus, we define the multipartite monogamy of coherence with respect to a measure as: 
\begin{equation}
M = \sum_{n=2}^{N} \mathcal{C}_{1:n} - \mathcal{C}_{1:2\dots N} ,
\label{q-def}
\end{equation}
which is monogamous for $ M \le 0 $ due to the multipartite coherence that is present.  For $M>0$ it is 
polygamous since the dominant coherence is distributed in a pairwise fashion.  

In Fig. \ref{fig2}(c) we calculate (\ref{q-def}) for the W states.  We find that $ M\ge 0 $ for all $\theta,\phi $, hence the state is strictly polygamous.  For the GHZ states as shown in Fig. \ref{fig2}(b) there is only one coherence contribution with $ \mathcal{C}_{1:n} = 0 $, which results in $ M = - \mathcal{C} $, meaning that it is strictly monogamous. This is as expected since the GHZ states are tripartite
entangled, whereas the $W$ state has a bipartite nature \cite{wd00}.  
For the Heisenberg spin chain we find both monogamous ($\Delta > 2.9$) and polygamous behavior ($-1 < \Delta < 2.9$) (see Fig. \ref{fig2}(d) inset).  For $\Delta \gg 1$ region when the ground state
is a N\'{e}el state, where the two-site coherences vanish $ \mathcal{C}_{1:n} \rightarrow 0 $.  Then the coherence is entirely due to the 
$1$:$2\dots N $ bipartition, resulting in a monogamous state. This can be understood to be due to the fact that the N\'{e}el state superposition is essentially the same as a GHZ state up to a redefinition of state labels.  For small $ \Delta $, there is a larger effect from the off-diagonal terms $ \sigma^x_n \sigma^x_{n+1} + \sigma^y_n \sigma^y_{n+1} = 2( \sigma^+_n \sigma^-_{n+1} + \sigma^-_n \sigma^+_{n+1}) $.  This term tends to create coherence on nearby sites, which is more characteristic of a polygamous behavior. In this way the 
parameter $ \Delta $ switches the nature of the coherence between monogamy and polygamy by redistribution it between relatively 
local sites to a genuinely multipartite form.

{\it Conclusions.}
Multipartite coherence is decomposed into local and intrinsic parts and quantified using a entropic
measure with metric nature. This decomposition into various contributions can be used not only to 
characterize a given state but also to locate the origin of the coherence.  In many cases there is a crossover behavior between the coherences
of different origins, which depends upon the type of the state examined.  In the transverse Ising model, the coherence transitions between 
local coherence on the sites to a GHZ-type multipartite nature.  The coherence decompositions leads to a multipartite monogamy inequality 
for coherence measures, giving another way of characterizing the nature of coherence in these systems.  In the Heisenberg $XXZ$ model the 
coherence displays a crossover between monogamous and polygamous behavior when the anisotropy parameter 
is varied.  The framework provided in this paper allows for a simple 
way to understand the nature of an arbitrary quantum state, by characterizing the various coherence contributions, even for relatively complicated states in quantum many-body problems.

In addition to providing a framework for decomposing coherence, we believe that the general method is potentially applicable in several contexts.  In the field of quantum simulation and quantum computing it is often of interest to understand what kind of 
quantum state is generated, either to understand the nature of a many-body system \cite{buluta09} or for the purposes of benchmarking \cite{rb14,tx15}.  Finding the distribution of coherence provides a more illuminating way of understanding the nature of a quantum state.   
 One of the contributions which quantum information made to condensed matter physics is the introduction of entanglement as a quantity that can be used to characterize the state of a system \cite{osborne02}. It is an interesting question of whether particular types of coherence could be used to analyze similarly quantum phase transitions.  Furthermore, quantum limits to shareability (i.e. monogamy) of entanglement
is known to be related to frustration in many body systems \cite{ko01,mw04,af07,xm11,sg15}, and affect the coherence and entanglement structure in the system.  This has a direct effect on approaches to efficiently capture the wavefunction of interacting quantum many-body systems, such as matrix product states and their variants \cite{fv06,gv08}.  In quantum metrology, a recent development has been the use of local rather than global strategies to gain interferometric advantages \cite{bh07,js15,pk16}, which highlights resource nature of coherence. 
An interesting future possibility for the QJSD is that due to its distance and metric properties and entropic nature, it could contribute to differential geometry based approaches to quantum information theory to understanding of the geometry of quantum states \cite{ib06}.

This work is supported by the Shanghai Research Challenge Fund, New York University Global Seed Grants for Collaborative
Research, National Natural Science Foundation of China grant 61571301, and the Thousand Talents Program for 
Distinguished Young Scholars.


\begin{thebibliography}{99}

\bibitem{vs15} V.S. Narasimhachar and G. Gour, Nature Communications 6, 7689 (2015).
%
\bibitem{ja14} J. Aberg, Phys. Rev. Lett 113, 150402 (2014).
%
\bibitem{ml15} M. Lostaglio, D. Jennings and T. Rudolph, Nature Communications 6, 6383 (2015). 
%
\bibitem{lo15} M. Lostaglio, K.Korzekwa, D. Jennings and T. Rudolph, Phy. Rev. X 5, 021001 (2015). 
%
\bibitem{ok11} O. Karlstrom, H. Linke, G. Kralstrom, and A. Wacker, Phys. Rev. B 84, 113415 (2011). 
%
\bibitem{ps07} G.S. Engel, Nature 446, 782 (2007), 
%
\bibitem{er14} E. Romero et. al, Nature Physics 10, 676 (2014). 
%
\bibitem{mn00} M.A. Nielsen and I.L. Chuang, {\it Quantum Computation and Quantum Information}
(Cambridge University Press, Cambridge, 2000). 
%
\bibitem{rg63} R.J. Glauber, Phys. Rev. 131, 2766 (1963).
%
\bibitem{es63} E.C.G Sudarshan, Phys. Rev. Lett. 10, 277 (1963). 
%
\bibitem{ms97} M.O. Scully and M.S. Zubairy, {\it Quantum Optics} (Cambridge University Press, Cambridge, 1997).
%
\bibitem{tb14} T. Baumgratz, M. Cramer and M.B. Plenio, Phys. Rev. Lett. 113, 140401 (2014).
%
\bibitem{dg14} D. Girolami, Phys. Rev. Lett 113, 170401 (2014).
%
\bibitem{dpp15} D.P. Pires, L.C. Celeri, D.O. Soares-Pinto, Phys. Rev. A 91, 042330 (2015). 
%
\bibitem{as15} A. Streltsov, U. Singh, H.S. Dhar, M.N. Bera and G. Adesso, Phys. Rev. Lett 115, 020403 (2015).
%
\bibitem{yy15} Y. Yao, X. Xiao, L. Ge and C.P. Sun, Phys. Rev. A 92, 022112 (2015).
%
\bibitem{vc00} V. Coffman, J. Kundu and W.K. Wootters, Phys. Rev. A 61, 052306 (2000).
%
\bibitem{mk04} M. Koashi and A. Winter, Phys. Rev. A 69, 022309 (2004).
%
\bibitem{to06} T.J. Osborne and F. Verstraete, Phys. Rev. Lett. 96, 220503 (2006).
%
\bibitem{ga06} G. Adesso and F. Illuminati, New. J. Phys. 8, 15 (2006). 
%
\bibitem{pk14} P. Kurzy\'nski, A. Cabello and D. Kaszlilkowski, Phys. Rev. Lett. 112, 100401, (2014).
%
\bibitem{vv97} V. Vedral, M.B. Plenio, M.A. Rippin and P.L. Knight, Phys. Rev. Lett 78, 2275 (1997).
%
\bibitem{vvp97} V. Vedral, M.B. Plenio, K. Jacobs and P.L. Knight, Phys. Rev. A 56, 4452 (1997).
%
\bibitem{vv98} V. Vedral and M.B. Plenio, Phys. Rev. A 57, 1619 (1998). 
%
\bibitem{jp09} J. Briet and P. Harremoes, Phys. Rev. A 79, 052311 (2009). 
%
\bibitem{ap05} A.P. Majtey, P.W. Lamberti and D.P. Prato, Phys. Rev. A 72 052310 (2005). 
%
\bibitem{pl08} P.W. Lamberti, A.P. Majtey, A. Borras, M. Casas and A. Plastino,
Phys. Rev. A 77 052311 (2008).
%
\bibitem{wd00} W. D\"{u}r, G. Vidal and J.I. Cirac, Phys. Rev. A 62, 062314 (2000).  
%
\bibitem{buluta09} I. Buluta and F. Nori, Science 326, 108 (2009).
%
\bibitem{rb14} R. Barends et. al. Nature 508, 500 (2014).  
%
\bibitem{tx15} T. Xia et al. Phys. Rev. Lett 114, 100503 (2015). 
%
\bibitem{osborne02}	T. J. Osborne and M. A. Nielsen, Phys. Rev. A 66, 032110 (2002).
%
\bibitem{ko01} K.M. O'Connor and W.K. Wootters, Phys. Rev. A 63, 052302 (2001).
%
\bibitem{mw04} M.M. Wolf, F. Verstraete and J.I. Cirac, Phys. Rev. Lett. 92, 087903 (2004).
%
\bibitem{af07} A. Ferraro, A. Garc\'{i}a-Saez and A. Ac\'{i}n, Phys. Rev. A 76, 052321 (2007). 
%
\bibitem{xm11} X.S. Ma, B. Dakic, W. Naylor, A. Zeilinger and P. Walther, Nature Physics 7, 399 (2011). 
%
\bibitem{sg15}	S.M. Giampaolo, B.C. Hiesmayr and F. Illuminati, Phys. Rev. B 92, 144406 (2015).
%
%\bibitem{fv04} F. Verstraete and J.I. Cirac,
%              {\it Renormalization algorithms for Quantum many-body systems in two and higher dimensions}
%               eprint arXiv No: cond-mat/0407066.
%
\bibitem{fv06} F. Verstraete and J.I. Cirac, Phys. Rev. B 73, 094423(2006).
%
\bibitem{gv08} G. Vidal, Phys. Rev. Lett. 101, 110501 (2008). 
%
%\bibitem{ge09} G. Evenbly and G. Vidal, Phys. Rev. B 79, 144108 (2009).
%
%\bibitem{jg12} J.P. Gaebler et al. Phys. Rev. Lett 108, 260503 (2012).
%
\bibitem{bh07} B.L. Higgins, D.W. Barry, S.D. Bartlett, H.M. Wiseman and G.J. Pyrde, Nature 450, 393 (2007).
%
\bibitem{js15}	J. Sahota and N. Quesada, Phys. Rev. A 91, 013808 (2015).
%
\bibitem{pk16} 	P.A. Knott, T.J. Proctor, A.J. Hayes, J.F. Ralph, P.Kok and J.A. Dunningham,
{\it Local versus Global strategies in multi-parameter estimation} eprint arXiv No: 1601.05912  [quant-ph].
%
\bibitem{ib06} I. Bengtsson and K. \.{Z}yckowski, {\it Geometry of Quantum States} (Cambridge University Press, Cambridge, 2006).
\end{thebibliography}
\end{document}